\documentclass[twocolumn,pre,superscriptaddress]{revtex4}

\usepackage{graphicx}
\usepackage{amsmath,amssymb}
\usepackage{subfigure}

\newcommand\defn{\textit}

\newcommand\e{\mathrm{e}}

\newcommand\av[1]{\langle#1\rangle}

\newcommand\half{\mbox{$\frac12$}}
\newcommand{\Ord}{\mathrm{O}}
\newcommand{\Rone}{\mathrm{R}_1}

\newcommand{\Ronene}{\mathrm{R}_1\textrm{-}\overline{\mbox{ext}}}
\newcommand{\Roneint}{\mbox{$\mathrm{R}_1$-int}}
\newcommand{\Rtwo}{\mathrm{R}_2}

\begin{document}

\title{Interacting epidemics and coinfection on contact networks}

\author{M. E. J. Newman}
\affiliation{Center for the Study of Complex Systems and Department of
  Physics, University of Michigan, Ann Arbor, MI 48109, U.S.A.}
\author{C. R. Ferrario}
\affiliation{Department of Pharmacology, University of Michigan, Ann Arbor,
MI 48109, U.S.A.}

\begin{abstract}
  The spread of certain diseases can be promoted, in some cases
  substantially, by prior infection with another disease.  One example is
  that of HIV, whose immunosuppressant effects significantly increase the
  chances of infection with other pathogens.  Such coinfection processes,
  when combined with nontrivial structure in the contact networks over
  which diseases spread, can lead to complex patterns of epidemiological
  behavior.  Here we consider a mathematical model of two diseases
  spreading through a single population, where infection with one disease
  is dependent on prior infection with the other.  We solve exactly for the
  sizes of the outbreaks of both diseases in the limit of large population
  size, along with the complete phase diagram of the system.  Among other
  things, we use our model to demonstrate how diseases can be controlled
  not only by reducing the rate of their spread, but also by reducing the
  spread of other infections upon which they depend.
\end{abstract}

\pacs{89.75.Hc,02.10.Ox,02.50.-r}

\maketitle

\section{Introduction}
\label{sec:intro}
Two diseases circulating in the same population of hosts can interact in
various ways.  One disease can, for instance, impart cross-immunity to the
other, meaning that an individual infected with the first disease becomes
partially or fully immune to infection with the
second~\cite{CHL96,Newman05c}.  A contrasting case occurs when infection
with one disease \emph{increases} the chance of infection with a second.  A
well-documented example is HIV, which, because of its immunosuppressant
effects, increases the chances of infection with a wide variety of
additional pathogens.  Other examples include syphilis and HSV--2, the
presence of either of which can substantially increase the chances of
contracting, for example, HIV~\cite{LL04,Freeman06,Perre08,Sartori11}.  In
a non-disease context, similar phenomena also arise in the epidemic-like
spread of fashions, fads, or ideas through a population.  There are, for
instance, many examples of products whose adoption or purchase depends on
the consumer already having adopted or purchased another product.  The
purchase of software or apps for computers or phones, for instance,
requires that the purchaser already own a suitable computer or phone.  In
cases where adoption of products spreads virally, by person-to-person
recommendation, a ``coinfection'' model of adoption may then be
appropriate.

In this paper we study mathematically the behavior of infections that
promote or are promoted by other infections in this way.  We consider a
model coinfection system with two diseases, both displaying
susceptible--infective--recovered (SIR) dynamics~\cite{AM91,Hethcote00}, in
which any individual may contract the first disease if exposed to it, but
the second disease can be contracted only by an individual previously
infected with the first.  This is a simplification of the more general
situation in which absence of the first disease decreases the chance of
infection with the second but does not eliminate it altogether.  It is,
however, a useful simplification, retaining many qualitative features of
the more general case, while also allowing us to solve for properties of
the model exactly.  Following previous work on competing
pathogens~\cite{Newman05c}, we assume the spread of our two diseases to be
well temporally separated, the first disease passing completely through the
population before the second one strikes, although arguments
of~\cite{KN11b} suggest that this assumption could be relaxed without
significantly altering the results.

The choice of SIR dynamics for our model appears at first to be less
appropriate for a disease like HIV, from which sufferers do not normally
recover.  However, HIV is mainly infective during its primary stage---the
first few weeks of infection---after which it enters an asymptomatic stage
where probability of transmission is much lower~\cite{Wawer05,HAF08}.  The
``recovered'' state of our model can mimic this behavior quite well, at
least for some populations with HIV.

Following the description outlined above one can easily write down a
fully-mixed compartmental model of our interacting diseases in the style of
traditional mathematical epidemiology, but the results are essentially
trivial.  The first disease spreads through the population according to
ordinary SIR dynamics, then the second spreads in the subset of individuals
infected by the first, but otherwise again following ordinary SIR dynamics.
No qualitatively new behaviors emerge.

Real diseases, however, are not fully mixed.  Rather, they spread over a
network of physical contacts between individuals, whose structure is known
to have a substantial impact on patterns of
infection~\cite{PV01a,Newman02c,CBBV06}.  As we will demonstrate, the
spread of our two interacting diseases shows a number of interesting
behaviors once the presence of such an underlying contact network is taken
into account.

\section{The model}
We study a network-based model of interacting pathogens spreading through a
single population, which we solve exactly using the cavity method of
statistical physics.  From our solution we are able to calculate the
expected number of individuals infected with each of the two diseases as a
function of disease parameters, as well as the epidemic thresholds and
complete phase diagram of the system.

Our model consists of a network of $n$ nodes, representing the individuals
in the modeled population, connected in pairs by edges representing their
contacts.  The spread of the first disease through the network is
represented by an SIR process in which all individuals start in the
susceptible (S) state except for a single individual who is in the
infective (I) state---the initial carrier of the first disease.  Infectives
recover after a certain time~$\tau$, which we take to be constant, but
while infective they have a fixed probability~$\beta$ per unit time of
passing the disease to their susceptible neighbors.  The probability of the
disease being transmitted in a short interval of time~$\delta t$ is thus
$\beta\,\delta t$ and the probability of it not being transmitted is
$1-\beta\,\delta t$.  Thus the probability of not being transmitted during
the entire time interval~$\tau$ is
\begin{equation}
\lim_{\delta t\to0} ( 1 - \beta\,\delta t )^{\tau/\delta t}
    = \e^{-\beta\tau},
\end{equation}
and the total probability of being transmitted, the so-called infectivity
or transmissibility~$T_1$ for the first disease, is
\begin{equation}
T_1 = 1 - \e^{-\beta\tau}.
\end{equation}
We will consider this quantity to be an input parameter to our theory.

Once the first disease has passed through the population, leaving every
member of the population in either the susceptible or the recovered state
with no infectives remaining, then the second disease starts to spread, but
with the important caveat that it can spread only among those who have
previously contracted, and then recovered from, the first disease, a state
that we will denote~$\Rone$.  The second disease spreads among these
individuals again according to an SIR process, and we will explicitly allow
for the possibility that the second disease has a different
transmissibility~$T_2$ from the first.  Note however that the second
disease is still transmitted over the same contact network as the first,
which can lead to nontrivial correlations between the probabilities of
infection with the two diseases.  Because the network is assumed the same
for both diseases our model is primarily applicable to pairs of diseases
having the same mode of transmission---two airborne diseases, for example,
or two sexually transmitted diseases.

When the second disease has passed entirely through the system, every
member of the population is left in one of three states: susceptible~(S),
meaning they have never contracted either disease; infected by and
recovered from the first disease, but uninfected by the second~(denoted
$\Rone$); or infected by and recovered from both diseases~($\Rtwo$).  Note
that there are no individuals who contract the second disease but not the
first, since the first is a necessary condition for infection with the
second.  The number of individuals in the $\Rone$ and $\Rtwo$ states tell
us the total number who contracted each of the two diseases, and hence the
size of the two outbreaks.  As we will see, there is a nontrivial phase
diagram describing the variation of these numbers with the
transmissibilities~$T_1$ and $T_2$ of the diseases.

To fully define our model we need also to specify the structure of the
network of contacts over which the diseases spread.  Many choices are
possible, including model networks or networks based on empirical data for
real contacts.  In this paper, we employ one of the most widely used model
networks as the substrate for our calculations, the so-called configuration
model~\cite{MR95,NSW01}.  The configuration model is a random graph model
in which the degrees of nodes---the number of connections they have to
other nodes---are free parameters that may be chosen from any distribution.
Numerous studies in recent years have shown the degrees of nodes to have a
large impact on the structure and behavior of networked
systems~\cite{AB02,Newman03d,Boccaletti06}, so a model that does not allow
for varying degrees would be missing one of the most important of network
properties.  In respects other than this, however, the configuration model
assumes random connections between nodes, which, it turns out, makes the
network simple enough that we can solve exactly for the behavior of our
two-disease system upon it.

The configuration model is completely specified by giving the number~$n$ of
nodes in the model network, which we will assume to be large, and the
probability distribution of the degrees.  The latter is parametrized by the
fraction~$p_k$ of nodes that have degree~$k$, for $k=0\ldots\infty$.  For
instance, one might specify the degrees to have a Poisson distribution:
\begin{equation}
p_k = \e^{-c}\,{c^k\over k!},
\label{eq:poisson}
\end{equation}
where $c$ is the average degree in the network as a whole.

An alternative way of thinking about~$p_k$ is as the probability that a
randomly chosen node has degree~$k$.  In our calculations we will also need
to consider randomly chosen edges and ask what the probability is that the
node at one end of such an edge has degree~$k$.  It is clear that this
probability cannot in general be equal to~$p_k$.  For instance, there is no
way to follow an edge and reach a node of degree zero, even if degree-zero
nodes exist in the network.  So nodes at the end of an edge must have some
other distribution of degrees.  In fact, the relevant quantity for the
purposes of this paper will be not the degree of the node at the end of an
edge, but the degree minus one, which is the number of edges attached to
the node other than the edge we followed to reach it.  This number, often
called the \defn{excess degree}, has distribution
\begin{equation}
q_k = {(k+1)p_{k+1}\over\av{k}},
\label{eq:qk}
\end{equation}
where $\av{k} = \sum_k k p_k$ is the average degree in the
network~\cite{Newman03d}.  The quantity~$q_k$ is called the \defn{excess
  degree distribution} and both $p_k$ and $q_k$ will play important roles
in the developments here.

Because they will be useful later, we also define probability generating
functions for the two distributions:
\begin{equation}
g_0(z) = \sum_{k=0}^\infty p_k z^k,\qquad
g_1(z) = \sum_{k=0}^\infty q_k z^k.
\label{eq:pgfs}
\end{equation}
In what follows, we will assume we know the degree distribution~$p_k$ of
our network, and hence that we know also the excess degree distribution,
from Eq.~\eqref{eq:qk}, and the two generating functions,
Eq.~\eqref{eq:pgfs}.

\section{Solution for the number of individuals infected}
We can solve exactly for the expected number of individuals infected by our
two diseases on configuration model networks with arbitrary degree
distributions.  The calculation for the first disease is the simpler of the
two, so we start there.  This part of the solution follows closely the
outline of our previous presentations in~\cite{CNSW00,Newman02c}.

\begin{figure}
\begin{center}
\includegraphics[width=4.0cm]{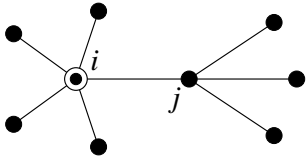}
\end{center}
\caption{We calculate the probability of infection of a node~$i$ (circled)
  with either one or both of the two diseases by first calculating the
  probability that a neighbor~$j$ is infected.  We must account separately
  for cases in which $j$ caught the first disease from $i$ itself or from
  another of its neighbors, since these two cases have different
  implications for the spread of the second disease.}
\label{fig:cavity}
\end{figure}

Consider Fig.~\ref{fig:cavity}, which depicts the neighborhood of a typical
node~$i$ somewhere in the network, and let us calculate the average
probability~$S_1$ that such a node will ever be infected by disease~1.  To
do this we first consider the probability that a neighbor of~$i$, call it
node~$j$, will be infected by disease~1 if $i$ is removed from the network.
Let us denote this latter probability by~$u$.

The removal of node~$i$ is a crucial element of our calculation.  Some
neighbors of~$i$ may be infected by $i$ itself, but such a neighbor cannot
then infect~$i$ back, since $i$ by definition already has the disease.
Thus in calculating the probability of $i$'s infection we need to discount
such processes and count only neighbors of~$i$ who were ``externally
infected,'' meaning they were infected by one of their neighbors other
than~$i$.  A simple way to achieve this is to remove $i$ entirely from the
network.

Once $i$ is removed from the network, the infection states of $i$'s
neighbors become statistically independent---the infection of one makes the
infection of others no more or less likely.  This is a particular property
of configuration model networks in the limit of large network size.  Such
networks contain closed loops of edges that could in principle induce
correlations between the states of nodes, but in the limit of large size
the length of these loops diverges and the correlations vanish.  The
statistical independence between the neighbors of $i$ is the crucial
property that makes exact calculations possible for our model.

If we know the value of~$u$, the probability of external infection of a
neighboring node of~$i$, then the value of~$S_1$, the average probability
of infection of~$i$ itself, is readily calculated as follows.  A
neighbor~$j$ of node~$i$ is infected with probability~$u$ and transmits
that infection to~$i$ with probability equal to the transmissibility~$T_1$,
for an overall probability of infection~$uT_1$.  Then the probability of
$i$ not being infected by~$j$ is $1-uT_1$ and the probability of $i$ not
being infected by any of its neighbors is $(1-uT_1)^k$ if it has
exactly~$k$ neighbors---the statistical independence of the neighbor states
means that the probability for all $k$ neighbors is just the probability
for a single one to the $k$th power.  Now averaging this quantity over the
degree distribution~$p_k$, we find the mean probability that $i$ is not
infected to be
\begin{equation}
\sum_{k=0}^\infty p_k (1-uT_1)^k = g_0(1 - uT_1),
\end{equation}
where $g_0$ is the generating function for the degree distribution defined
in Eq.~\eqref{eq:pgfs}.  Then the probability that $i$ is infected is
\begin{equation}
S_1 = 1 - g_0(1 - uT_1).
\label{eq:S1}
\end{equation}

It remains for us to find the value of~$u$, which we can do by an analogous
calculation.  The probability that neighboring node~$j$ is not (externally)
infected takes the form $(1-uT_1)^k$, just as for node~$i$, except that $k$
now represents the number of external neighbors of~$j$, neighbors other
than~$i$.  This is the number we previously called the excess degree
of~$j$, and it is distributed according to the excess degree distribution
of Eq.~\eqref{eq:qk}.  Averaging over this distribution, the mean
probability that $j$ (or any neighbor node) is not infected is given by
\begin{equation}
\sum_{k=0}^\infty q_k (1-uT_1)^k = g_1(1 - uT_1),
\end{equation}
where $g_1$ is the generating function for the excess degree distribution.
Then the probability that $j$ is externally infected is
\begin{equation}
u = 1 - g_1(1 - uT_1).
\label{eq:u}
\end{equation}

Between them, Eqs.~\eqref{eq:S1} and~\eqref{eq:u} allow us to solve for the
average probability of infection of a node by disease~1: we first solve the
self-consistent condition~\eqref{eq:u} for the value of~$u$, then we
substitute the result into~\eqref{eq:S1} to get the value of~$S_1$.  Note,
moreover, that if $S_1$ is the probability of infection, then $nS_1$ is the
expected number of individuals infected with disease~1, so this calculation
also gives us the expected size of the outbreak of disease~1.

This calculation is an example of a \defn{cavity method}, a~technique
commonly used in statistical physics for the solution of network and
lattice problems.  The word ``cavity'' refers to the node~$i$ which is
removed, leaving a hole or cavity in the network.  The calculation above is
a particularly simple example of the cavity method.  The calculation of the
spread of the second disease, however, which also makes use of the cavity
method, is less simple.

Consider then the equivalent calculation for the second disease, in which
we calculate the average probability that a node is infected with the
second disease, the first disease having already spread through the system.
An important point to recognize is that the subset of nodes through which
the second disease spreads, which is the subset that was previously
infected with the first disease, does not itself form a configuration model
network.  This is made clear for instance by the fact that the subset in
question is connected---it forms a single network component---which is not
true in general of configuration model networks~\cite{NSW01}.  As a
result, our cavity method calculation is not the same for disease~2 as it
was for disease~1, being rather more delicate.  In particular, as we will
see, the cavity node~$i$ must now be removed for some parts of the
calculation but not for others.  We will break the calculation down into a
number of steps.

First, when disease~1 spreads, node~$i$ either gets infected (with
probability $S_1$ given by Eq.~\eqref{eq:S1} above) or it does not (with
probability $1-S_1$).  If it is not infected then it cannot later be
infected with disease~2, and hence our calculation is finished---we need go
no further.  In all subsequent steps, therefore, we will assume that $i$
has been infected with (and has then recovered from) disease~1, a state
that we previously denoted~$\Rone$.

Suppose that node~$i$ has degree~$k$.  Let us ask what the value is of the
probability~$P(\Rone,m|k)$ that it was infected with disease~1 and also has
exactly $m$ neighbors who were externally infected with disease~1, meaning
that they were infected by any of \emph{their} neighbors other
than~$i$---see Fig.~\ref{fig:cavity} again.  We note that $i$ must have
contracted disease~1 from one of its externally infected neighbors and the
probability of this happening is
\begin{equation}
P(\Rone|m) = 1 - (1-T_1)^{m}.
\end{equation}
Also, since the probability of a neighbor's external infection with
disease~1 is~$u$ by definition, the probability of having $m$ externally
infected neighbors is
\begin{equation}
P(m|k) = {k\choose m} u^m (1-u)^{k-m}.
\end{equation}
Combining these expressions, we have
\begin{align}
P(\Rone,m|k) &= P(\Rone|m)\,P(m|k) \nonumber\\
  &= {k\choose m} u^{m} (1-u)^{k-m} \bigl[ 1 - (1-T_1)^{m} \bigr].
\label{eq:Pm1}
\end{align}

The number~$m$, however, does not reflect the total number of $i$'s
neighbors who have had disease~1 because, in addition to those infected
externally as above, some number~$m'$ of the $k-m$ remaining nodes may also
have been infected directly by~$i$ itself.  (This is the part of the
calculation in which $i$ must not be considered removed from the network.)
Given that $i$ has had disease~1, the probability of such a direct
infection for a neighbor of $i$ is just $T_1$ and hence
\begin{equation}
P(m'|\Rone,m,k) = {k-m\choose m'} T_1^{m'} (1-T_1)^{k-m-m'}.
\label{eq:Pm1p}
\end{equation}

\begin{widetext}
Combining Eqs.~\eqref{eq:Pm1} and~\eqref{eq:Pm1p} we have
\begin{equation}
P(\Rone,m,m'|k) = P(m'|\Rone,m,k)\,P(\Rone,m|k)
  = {k\choose m} u^{m} (1-u)^{k-m} \bigl[ 1 - (1-T_1)^{m} \bigr]
    {k-m\choose m'} T_1^{m'} (1-T_1)^{k-m-m'}.
\end{equation}
And, multiplying by the probability~$p_k$ of having degree~$k$, summing
over~$k$, then dividing by the prior probability $P(\Rone) = S_1$ of
contracting disease~1, we get
\begin{equation}
P(m,m'|\Rone) = {1\over S_1} \sum_{k=0}^\infty p_k
                {k\choose m} u^{m} (1-u)^{k-m}
                \bigl[ 1 - (1-T_1)^{m} \bigr]
                {k-m\choose m'} T_1^{m'} (1-T_1)^{k-m-m'}.
\label{eq:PmmRone}
\end{equation}
This quantity represents the probability that a node~$i$ that has had
disease~1 has $m+m'$ neighbors who have also had disease~1, of whom $m'$
were infected by~$i$ itself and the remaining~$m$ contracted their
infections from other sources.

We can usefully encapsulate this rather complicated expression in a double
generating function~$h_0(y,z)$ for the number of infected neighbors of~$i$
thus:
\begin{align}
h_0(y,z) &= \sum_{m,m'} P(m,m'|\Rone)\,y^m {z}^{m'} \nonumber\\
  &= {1\over S_1}\sum_{k=0}^\infty p_k\!\! \sum_{m=0}^k \!
     {k\choose m} u^{m} (1-u)^{k-m} \bigl[ 1 - (1-T_1)^{m} \bigr]
     y^m \sum_{m'=0}^{m} {k-m\choose m'} T_1^{m'} (1-T_1)^{k-m-m'} {z}^{m'}
   \nonumber\\
  &= {1\over S_1} \sum_{k=0}^\infty p_k \Bigl(
   \bigl[ u y + (1-u)(1-T_1+zT_1) \bigr]^k
    - \bigl[ u(1-T_1)y + (1-u)(1-T_1+zT_1) \bigr]^k \Bigr) \nonumber\\
  &= {1\over S_1} \Bigl( g_0\bigl[ u y + (1-u)(1-T_1+zT_1) \bigr]
   - g_0\bigl[ u(1-T_1)y + (1-u)(1-T_1+zT_1) \bigr] \Bigr),
\label{eq:h0}
\end{align}
where $g_0(z)$ is the generating function for the degree distribution
defined in Eq.~\eqref{eq:pgfs}.  (As a check on this formula, we note that
if we set $y=z=1$ we should get $h_0(1,1) = 1$.  We leave it as an exercise
for the particularly avid reader to demonstrate that this is indeed true.)
\end{widetext}

Given these results, the probability~$S_2$ that node~$i$ is infected with
disease~2 given that it was previously infected with disease~1, is
calculated as follows.  Let $v$ be the probability that a neighbor of
node~$i$ is externally infected with disease~2 (i.e.,~not via node~$i$)
given that it has already been externally infected with disease~1.  Then
the probability that $i$ is infected with disease~2 by a neighbor that
externally contracted disease~1 is $vT_2$ and if there are $m$ such
neighbors in total then the probability of $i$ not contracting disease~2
from any of them is $(1-vT_2)^m$.

Conversely, let $w$ be the probability that a neighbor of~$i$ is externally
infected with disease~2 given that it was \emph{internally} infected with
disease~1, meaning it was infected directly by node~$i$.  (As we will see
in a moment, the probabilities~$v$ and $w$ are not the same, so we must
treat them separately.)  Then the probability that $i$ fails to contract
disease~2 from any of the $m'$ such nodes is $(1-wT_2)^{m'}$.

Combining these results, the probability that $i$ does not contract
disease~2 at all is $(1-vT_2)^m (1-wT_2)^{m'}$ and the probability that it
does is one minus this quantity.  Averaging over~$m$ and~$m'$, we find that
\begin{align}
S_2 &= P(\Rtwo|\Rone) \nonumber\\
  &= \sum_{m,m'} P(m,m'|\Rone) \bigl[ 1- (1-vT_2)^m
     (1-wT_2)^{m'}\bigr] \nonumber\\
  &= 1 - h_0(1-vT_2,1-wT_2),
\label{eq:S2}
\end{align}
where we have made use of the generating function~$h_0$ defined in
Eq.~\eqref{eq:h0}.

This expression is the equivalent of Eq.~\eqref{eq:S1} for the probability
of infection with disease~2.  It gives us the mean probability that an
individual is infected with disease~2 given that it was previously infected
with disease~1.  Alternatively, $S_2$~is the fraction of those individuals
infected with disease~1 that also contract disease~2, $S_1S_2$~is the
fraction of individuals in the entire network that contract disease~2, and
$nS_1S_2$ is the expected number of individuals with disease~2.

We have yet to calculate the values of the quantities~$v$ and~$w$, but
these calculations are now quite straightforward.  The calculation of~$v$
is the exact analog of the calculation we have already performed for~$S_2$.
We calculate the probability that a neighbor of~$i$ itself has $m$
(or~$m'$) neighbors externally (internally) infected with disease~1, which
is given by Eq.~\eqref{eq:PmmRone} but with $S_1$ replaced with $u$ and
$p_k$ replaced with~$q_k$.  The generating function for this distribution
is then the natural generalization of Eq.~\eqref{eq:h0}:
\begin{align}
h_1(y,z) &= {1\over u} \Bigl[ g_1\bigl[ u y + (1-u)(1-T_1+zT_1) \bigr]
     \nonumber\\
  &\quad {} - g_1\bigl[ u(1-T_1)y + (1-u)(1-T_1+zT_1) \bigr] \Bigr].
\label{eq:h1}
\end{align}
Then $v$ is the solution to the self-consistent condition
\begin{equation}
v = 1 - h_1(1-vT_2,1-wT_2),
\label{eq:v}
\end{equation}
which is analogous to Eq.~\eqref{eq:S2}.

The calculation of~$w$ is a little trickier.  Recall that $w$ is the
probability that $i$'s neighbor~$j$ is externally infected with disease~2
given that it was \emph{internally} infected with disease~1 (i.e.,~via
node~$i$).  If $j$ has exactly $m$ neighbors (other than~$i$) that were
externally infected with disease~1, then the probability that all of them
failed to infect~$j$ is $P(\Ronene|m) = (1-T_1)^m$, where the notation
``$\Ronene$'' denotes that $j$ was not externally infected.  If $j$ has
excess degree~$k$ then $P(m|k)={k\choose m}u^m(1-u)^{k-m}$, and
\begin{equation}
P(\Ronene,m|k) = {k\choose m}u^m(1-u)^{k-m} (1-T_1)^m.
\label{eq:tr1}
\end{equation}
The number~$m'$ of neighbors of~$j$ infected with disease~1 by $j$ itself
is distributed according to
\begin{equation}
P(m'|\Roneint,m,k) = {k-m\choose m'} T_1^{m'} (1-T_1)^{k-m-m'},
\end{equation}
where ``$\Roneint$'' denotes that $j$ was internally infected.  Noting that
$P(\Roneint|\Ronene,m,k) = T_1$ since $i$ has presumptively had disease~1
and has probability~$T_1$ of having transmitted it to~$j$ regardless of the
values of $m$ and~$k$, we have
\begin{align}
& P(\Roneint,m'|\Ronene,m,k) = \nonumber\\
  &\hspace{2em} P(m'|\Roneint,m,k) P(\Roneint|\Ronene,m,k) \nonumber\\
  &\hspace{0.7em}{} = T_1 {k-m\choose m'} T_1^{m'} (1-T_1)^{k-m-m'}.
\label{eq:tr2}
\end{align}
Multiplying Eqs.~\eqref{eq:tr1} and~\eqref{eq:tr2} and noting that
$\Roneint$ always implies $\Ronene$, we get an expression for
$P(\Roneint,m,m'|k)$.  Then we multiply by~$q_k$ and sum over~$k$ to get
$P(\Roneint,m,m')$, and divide by the prior probability $(1-u)T_1$ of being
internally infected with disease~1 to get
\begin{align}
P(m,m'|\Roneint) &= {1\over1-u} 
                    {k\choose m}u^m(1-u)^{k-m} (1-T_1)^m \nonumber\\
  &\qquad {} \times {k-m\choose m'} T_1^{m'} (1-T_1)^{k-m-m'}.
\end{align}
The generating function for this probability distribution is
\begin{equation}
h_2(y,z) = {g_1\bigl[ u(1-T_1)y + (1-u)(1-T_1+zT_1) \bigr]\over 1-u}.
\label{eq:h2}
\end{equation}
Finally, $w$~itself is given by the equivalent of Eq.~\eqref{eq:v}:
\begin{equation}
w = 1 - h_2(1-vT_2,1-wT_2).
\label{eq:w}
\end{equation}

Our complete prescription for calculating the number of nodes infected with
both diseases is now as follows.  (1)~We solve Eqs.~\eqref{eq:S1}
and~\eqref{eq:u} for~$u$ and~$S_1$; (2)~we use the value of~$u$ to solve
Eqs.~\eqref{eq:v} and~\eqref{eq:w} for~$v$ and~$w$, given the definitions
of $h_1$ and $h_2$ in Eqs.~\eqref{eq:h1} and~\eqref{eq:h2}; (3)~we
substitute the resulting values into Eq.~\eqref{eq:S2} to find~$S_2$.

As an added bonus, the quantities~$S_1$ and $S_2$ also tell us the
probabilities of epidemic outbreaks of each of our two diseases.  As
discussed in Ref.~\cite{Newman02c}, not all outbreaks of a disease reach a
large fraction of the population.  The infection process is stochastic and
sometimes, by luck, a disease starting with a single initial carrier will
not get passed to anyone else, or will get passed to only a few and then
fizzle out.  Other times it will take off and become an epidemic, and the
probability of it doing this is exactly equal to the fraction of the
network ultimately infected with the disease.  Thus the probability of an
epidemic outbreak of disease~1 is simply~$S_1$, and the probability of an
epidemic outbreak of disease~2 is $S_2$ given that an outbreak of disease~1
already happened, or $S_1S_2$ overall.

\section{Epidemic thresholds}
It is possible for either $S_1$ or $S_2$ to be exactly zero, in which case
there will under no circumstances be an epidemic of the corresponding
disease.  In general there will be threshold values of the transmission
probabilities $T_1$ and $T_2$ below which no epidemics occur and we can
calculate the position of these epidemic thresholds from the equations
given in the previous section.

First consider disease~1, which is the simpler of the two.  The size of the
outbreak of disease~1 falls to zero when $u=0$, since this is the point at
which the probability of a node catching the disease from its network
neighbors vanishes.  (We can confirm this directly by setting $u=0$ in
Eq.~\eqref{eq:S1}, which gives $S_1=0$ since $g_0(1)=1$.)  The value of $u$
is given by Eq.~\eqref{eq:u}.  When we approach the epidemic transition
from above, $u$~becomes small and we can expand the equation in powers of
this small parameter as
\begin{equation}
u = 1 - g_1(1) + u T_1 g_1'(1) + \Ord(u^2),
\end{equation}
where the prime denotes differentiation.  But $g_1(1)=1$ and the
higher-order terms can be dropped in the limit as $u\to0$, and hence we
find the value of~$T_1$ in this limit, which is by definition the epidemic
threshold value~$T_1^*$, to be
\begin{equation}
T_1^* = {1\over g_1'(1)}.
\label{eq:T1star}
\end{equation}
This is a well known result which appears elsewhere in the
literature~\cite{Newman02c}.

For the second disease there are two ways in which the disease can fail to
create an epidemic.  The first is that disease~1 fails to create an
epidemic, in which case disease~2 must also fail, since it depends on
disease~1 for its propagation.  The second is that disease~1 creates an
epidemic, but the transmissibility of disease~2 is not high enough to
create a second epidemic among the subset of the population infected with
disease~1.  Assuming we are in this second regime, the size of the second
epidemic goes to zero when $v=w=0$ where $v$~and $w$ are the simultaneous
solutions of Eqs.~\eqref{eq:v} and~\eqref{eq:w}.  Applying the same method
as for disease~1, we consider a point slightly above the epidemic
threshold, where $v$ and~$w$ are small, and we expand in both to get
\begin{align}
v &= 1 - h_1(1,1) + v T_2 h_1^{(1,0)}(1,1) + w T_2 h_1^{(0,1)}(1,1) +
     \ldots, \nonumber\\
w &= 1 - h_2(1,1) + v T_2 h_2^{(1,0)}(1,1) + w T_2 h_2^{(0,1)}(1,1) +
     \ldots,
\end{align}
where the superscript $(a,b)$ denotes differentiation of the generating
functions with respect to their first and second arguments $a$ and $b$
times respectively.  Observing that $h_1(1,1) = h_2(1,1) = 1$ and
neglecting the higher terms in the limit as we go to the epidemic
transition, we have in matrix notation
\begin{equation}
\begin{pmatrix}
  h_1^{(1,0)} & h_1^{(0,1)} \\
  h_2^{(1,0)} & h_2^{(0,1)}
\end{pmatrix}
\begin{pmatrix}
  v_{\vphantom{1}}^{\vphantom{(}} \\ w_{\vphantom{1}}^{\vphantom{(}}
\end{pmatrix}
= {1\over T_2^*}
\begin{pmatrix}
  v_{\vphantom{1}}^{\vphantom{(}} \\ w_{\vphantom{1}}^{\vphantom{(}}
\end{pmatrix}.
\end{equation}
where the derivatives are evaluated at the point $(1,1)$.  In other words
$1/T_2^*$ is an eigenvalue of the $2\times2$ matrix on the left-hand side.

The eigenvalues of a general $2\times2$ matrix are equal to
$\half\bigl(\tau\pm\sqrt{\tau^2-4\Delta}\bigr)$, where $\tau$ and $\Delta$
are the trace and determinant of the matrix.  Making use of the definitions
of $h_1$ and~$h_2$ in Eqs.~\eqref{eq:h1} and~\eqref{eq:h2}, we find the
four derivatives appearing in our matrix to be
\begin{align}
h_1^{(1,0)}(1,1) &= g_1'(1) - (1-T_1) g_1'(1 - uT_1), \\
h_1^{(0,1)}(1,1) &= {1-u\over u} T_1 \bigl[ g_1'(1) - g_1'(1 - uT_1) \big], \\
h_2^{(1,0)}(1,1) &= {u\over1-u} (1-T_1) g_1'(1 - uT_1), \\
h_2^{(0,1)}(1,1) &= T_1 g_1'(1 - uT_1),
\end{align}
which means
\begin{align}
\label{eq:tau}
\tau &= g_1'(1) - (1-2T_1) g_1'(1 - uT_1), \\
\label{eq:Delta}
\Delta &= T_1^2 g_1'(1) g_1'(1-uT_1).
\end{align}
It remains only to determine which of the two eigenvalues gives the correct
result for~$T_2^*$.  This can be done by setting $T_1=1$, which gives
$\tau=g_1'(1)+g_1'(1-u)$ and $\Delta=g_1'(1)g_1'(1-u)$, and hence the two
eigenvalues are $g_1'(1)=1/T_1^*$ and $g_1'(1-u)$.  Logic dictates that the
first eigenvalue must be the correct choice: when $T_1=1$ the second
disease is spreading on the entire network and hence its epidemic threshold
must fall at $T_1^*$.  Thus, we find that
\begin{equation}
T_2^* = {2\over\tau+\sqrt{\tau^2-4\Delta}},
\label{eq:T2star}
\end{equation}
where $\tau$ and $\Delta$ are given by Eqs.~\eqref{eq:tau}
and~\eqref{eq:Delta}.

Notice that if we take the limit $T_1\to T_1^*=1/g_1'(1)$ from above, which
implies that $u\to0$, then we have $\tau=2$ and $\Delta=1$ and hence
$T_2^*=1$.  That is, when we are precisely at the epidemic threshold for
the first disease, the threshold for the second disease is~1.  We expect
$T_2^*$~to be a monotone decreasing (or at least non-increasing) function
of increasing~$T_1$ and when $T_1=1$ we have $T_2^*=T_1^*$ as shown above.
So we expect $T_2^*$ to be monotone decreasing in~$T_1$ and $T_1^*\le
T_2^*\le 1$ at all times.

Thus the epidemic threshold for disease~2 is never lower than the epidemic
threshold for disease~1.  The intuitive explanation of this result is that
the constraint on disease~2, that it spread solely among individuals
already infected with disease~1, only ever reduces the set of nodes it can
spread on and hence makes it harder, never easier, for the disease to
spread.

\section{Examples}
As a concrete example of the results of the previous sections, consider
interacting diseases spreading on a network with a Poisson degree
distribution as in Eq.~\eqref{eq:poisson}.  This distribution presents a
particularly simple case, because the excess degree distribution is equal
to the ordinary degree distribution $q_k=p_k$ and their two generating
functions are equal
\begin{equation}
g_0(z) = g_1(z) = \e^{c(z-1)}.
\end{equation}
Thus~$S_1=u$ and $u$ is a solution of
\begin{equation}
u = 1 - \e^{-cT_1u}.
\label{eq:poissonu}
\end{equation}
Similarly $S_2=v$ and $v$ and~$w$ are solutions of Eqs.~\eqref{eq:v}
and~\eqref{eq:w}, though neither of the latter equations is very simple.

The epidemic threshold for disease~1 in this case is
\begin{equation}
T_1^* = {1\over g_1'(1)} = {1\over c},
\label{eq:poissonT1}
\end{equation}
a well known result for a single disease on a Poisson random graph.  The
epidemic threshold for the second disease is given by
Eq.~\eqref{eq:T2star}.  Noting that $g_1'(z) = cg_1(z)$, the values
of~$\tau$ and $\Delta$ are
\begin{equation}
\tau = c u+2cT_1(1-u),
\quad\Delta = c^2 T_1^2 (1-u),
\end{equation}
which gives
\begin{equation}
T_2^* = {2\over c\bigl[ u+2T_1(1-u) + \sqrt{u^2 + 4T_1(1-T_1)u(1-u)}
         \bigl]}.
\label{eq:poissonT2}
\end{equation}

\begin{figure}
\begin{center}
\includegraphics[width=\columnwidth]{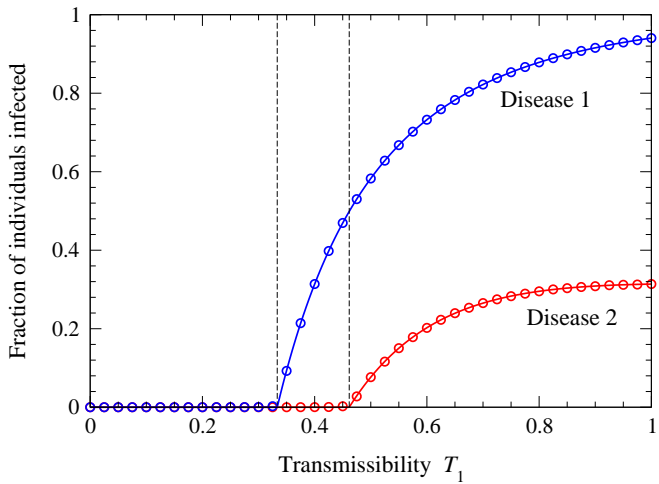}
\end{center}
\caption{The number of individuals infected with the two diseases on a
  network with a Poisson degree distribution with mean degree~$c=3$, as a
  function of the transmissibility~$T_1$ of the first disease.  The
  transmissibility of the second disease is fixed at $T_2=0.4$.  The solid
  curves show the analytical solutions, Eqs.~\eqref{eq:S1}
  and~\eqref{eq:S2}, while the points show the results of numerical
  simulations of the model.  Each point is an average of simulations on 100
  networks of a million nodes each.  Error bars are smaller than the points
  in all cases.  The two vertical dashed lines indicate the positions of
  the epidemic thresholds for the two diseases, from Eqs.~\eqref{eq:T1star}
  and~\eqref{eq:poissonT2}.}
\label{fig:poisson}
\end{figure}

Equations~\eqref{eq:v}, \eqref{eq:w}, and~\eqref{eq:poissonu} cannot be
solved exactly, but one can solve them by numerical iteration.  We choose
suitable starting values for $u$, $v$ and~$w$ (we find $u=v=w=\half$ to
work well) and iterate the equations to convergence.
Figure~\ref{fig:poisson} shows the resulting solutions for the sizes $S_1$
and $S_1S_2$ of the two disease outbreaks, as a function of~$T_1$ for a
network with average degree $c=3$ and a fixed value of $T_2=0.4$.  When
$T_1$ is small we are below the epidemic threshold $T_1^*=\frac13$ for the
first disease, marked by the first vertical line in the figure, and hence
neither disease spreads.  Above this point the first disease starts to
spread but does not, at least at first, infect enough individuals to allow
the spread of the second disease.  The system goes through a another
transition, marked by the second vertical line in the figure, when the size
of the first outbreak becomes large enough to support an outbreak of the
second.  This occurs at the value of~$T_1$ for which
Eq.~\eqref{eq:poissonT2} equals~$T_2$.

Thus, in this scenario, it would be possible to eradicate the second
disease by either one of two methods: one could take the traditional
approach of reducing its transmissibility~$T_2$ below the threshold
value~$T_2^*$, or, alternatively, one could reduce the transmissibility of
disease~1 until sufficiently few individuals are infected to allow the
spread of disease~2.

Also shown in the figure are numerical results from simulations of the
model on computer generated networks with the same Poisson degree
distribution.  As we can see, agreement between the analytic solution and
the numerical results is excellent.

\begin{figure}
\begin{center}
\includegraphics[width=8cm]{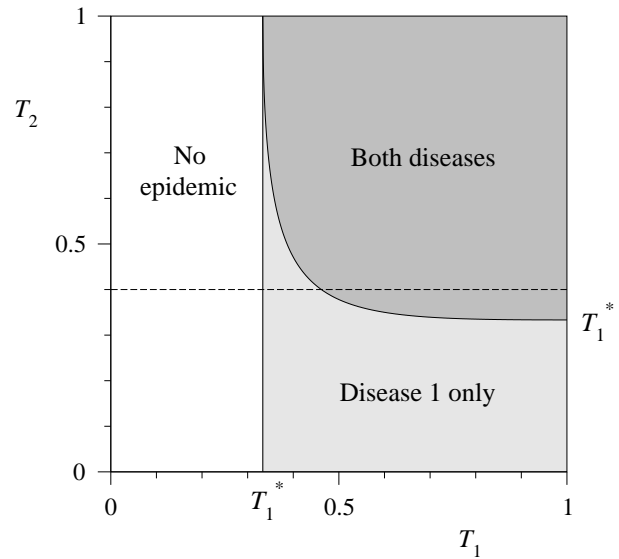}
\end{center}
\caption{Phase diagram of the model for a network with a Poisson degree
  distribution with mean degree $c=3$.  The horizontal dashed line
  represents the parameter values used for Fig.~\ref{fig:poisson}.}
\label{fig:phase}
\end{figure}

Using the values of $T_1^*$ and $T_2^*$ from Eqs.~\eqref{eq:poissonT1}
and~\eqref{eq:poissonT2} we can also plot a phase diagram for the model, as
in Fig.~\ref{fig:phase}, showing the regions in the $(T_1,T_2)$ parameter
space in which neither, one, or both of the diseases spread.  The
horizontal dashed line in the figure represents the parameter values used
in Fig.~\ref{fig:poisson}.

As another example, consider a network with a power-law degree
distribution.  As pointed out by Pastor-Satorras and
Vespignani~\cite{PV01a}, the epidemic threshold for a single disease on
such a network falls at $T_1^*=0$ provided the exponent of the power law is
less than~3.  This means that the disease always produces an epidemic
outbreak, no matter how low its transmissibility.  From the results above
we can show that the same will be true for both diseases in our two-disease
coinfection system.  The first disease behaves exactly as would a single
disease spreading on its own, and hence previous results such as those of
Ref.~\cite{PV01a} apply and $T_1^*=0$.  Alternatively, one can evaluate the
generating function $g_1(z)$ and show that $g_1'(1)\to\infty$ in a
power-law network and hence, by Eq.~\eqref{eq:T1star}, we have $T_1^*=0$.
Given that $g_1'(1)\to\infty$, however, we also see that $\tau\to\infty$
and $\Delta\to\infty$ in Eqs.~\eqref{eq:tau} and~\eqref{eq:Delta}, and
hence that $T_2^*=0$ in Eq.~\eqref{eq:T2star}.  In other words, the second
disease will also always spread, no matter how low the transmissibility of
either the first or second diseases.  In this case the second disease
cannot be eradicated by lowering either of the transmission probabilities.

The intuitive explanation of this result is that the subnetwork over which
the second disease spreads, which consists of those individuals infected
with the first disease, also has a power-law tail to its degree
distribution; the probability of infection with disease~1 increases with
node degree and tends to one in the limit of large degree, so that the
degree distribution of infected nodes is the same as that of the network as
a whole in the large-degree limit.  And it is only the power-law tail that
is needed to drive the epidemic threshold to zero---it is not required that
the distribution follow a pure power law over its entire domain.

Even though both diseases may spread, however, it is not necessarily the
case that many individuals are infected.  Indeed the number of individuals
infected with disease~1 will necessarily go to zero asymptotically as
$T_1\to0$, and hence so also will the number infected with disease~2 (which
can never exceed the number infected with disease~1).

\section{Conclusions}
In this paper we have studied a simple model of coinfection with two
diseases that spread over the same network of contacts.  In this model one
disease can spread freely through the population, limited only by its
probability of transmission, but the second disease can infect only those
infected with the first.  The result is a system displaying two distinct
epidemic thresholds, one occurring when the transmission probability of the
first disease reaches a high enough value to support an epidemic outbreak,
and the second occurring when the first disease infects a large enough
fraction of the population to allow spread of the second disease.  Thus,
while the first disease can (on a given network) be controlled only by
reducing its probability of transmission, the second can be controlled
either by reducing transmission or by reducing the number of individuals
infected with the first disease.

We have given an analytic solution for the size of both outbreaks and the
position of both thresholds on networks generated using the so-called
configuration model, for any choice of the degree distribution.  The
solution is exact in the limit of large network size and shows good
agreement with numerical simulations for large but finite networks.  We
have discussed two specific examples, of a network with a Poisson degree
distribution and a network with a power-law degree distribution.  In the
former case we find a distinct epidemic threshold for the second disease
that depends on the transmission probability for the first disease in such
a way that the second disease can be controlled or eradicated by reducing
either its probability of transition or that of the first disease.  In the
power-law case, by contrast, we find that the epidemic threshold for both
diseases falls at transmission probability zero, so that both will always
spread, no matter how low the transmission probabilities are.

A number of questions are unanswered by our analysis.  In particular, we
have not addressed any dynamical features of the epidemic process, such as
the time-scales or rate of growth of the epidemics.  And we have considered
only the case where the two diseases spread at well separated times.  If
they were to spread at the same time, it is possible one might see an
additional dynamical transition of the kind seen, for example,
in~\cite{KN11b}.

Furthermore our model covers only the case in which infection with the
first disease is a necessary condition for spread of the second, and not
the more general case where the first disease enhances transmission of the
second but is not an absolute requirement.  These issues, however, we leave
for future work.

\begin{acknowledgments}
  The authors thank Brian Karrer for useful conversations.  This work was
  funded in part by the National Science Foundation under grant
  DMS--1107796.
\end{acknowledgments}

\end{document}